\input psfig
\documentstyle{l-aa}     % LaTeX A&A  Standard Fonts
\newcommand{\sig}{\:\lower0.6ex\hbox{$\stackrel{\textstyle >}{\sim}$}\:}
\newcommand{\sil}{\:\lower0.6ex\hbox{$\stackrel{\textstyle <}{\sim}$}\:}
\newcommand{\sigs}{\:\lower0.4ex\hbox{$\stackrel{\scriptstyle >}{\scriptstyle \sim}$}\,}
\newcommand{\sils}{\:\lower0.4ex\hbox{$\stackrel{\scriptstyle <}{\scriptstyle \sim}$}\,}
\def\lsim{~\rlap{$<$}{\lower 1.0ex\hbox{$\sim$}}}
\def\gsim{~\rlap{$>$}{\lower 1.0ex\hbox{$\sim$}}}
\begin{document}
\thesaurus{02 % A&A Section 2: Cosmology
(11.03.1;     % Galaxies: clusters: general
12.03.1;      % {\it (Cosmology:)} cosmic microwave background
12.03.3;      % Cosmology: observations
12.04.2;      % {\it (Cosmology:)} diffuse radiation
13.25.3)      % X-rays: general
}

\title{Search for correlations between {\it COBE} DMR and {\it ROSAT} PSPC
All-Sky survey data}

%  \subtitle{}

\author{
	R\"udiger Kneissl 	\inst{1}
\and   Roland Egger		\inst{2} 
\and   G\"unther Hasinger	\inst{3} 
\and   Andrzej M. So\l tan	\inst{4} 
\and   Joachim Tr\"umper	\inst{2} 
}
\offprints{R. Kneissl,\\ ruk@mpa-garching.mpg.de}

\institute{Max-Planck-Institut f\"ur Astrophysik,
Karl-Schwarzschild-Str.1, D-85748 Garching bei 
M\"unchen, Germany 
\and Max-Planck-Institut f\"ur Extraterrestrische Physik, D-85748
Garching bei M\"unchen, Germany
\and Astrophysikalisches Institut, An der Sternwarte 16, D--14482
Potsdam, Germany
\and Copernicus Astronomical Center, Bartycka 18, 00-716 Warsaw, Poland
}

\date{Received July 10, accepted October 1, 1996}

\maketitle

\markboth{R. Kneissl et al.: \, Search for correlations between {\it COBE} DMR
and {\it ROSAT} PSPC All-Sky survey data}
         {R. Kneissl et al.: \, Search for correlations between {\it COBE} DMR
and {\it ROSAT} PSPC All-Sky survey data}

%\begin{center}
%\LARGE{A\&A -- revised}
%\end{center} 

\begin{abstract}
%_____________________________________ Do not leave a blank line here!
Results from a cross-correlation analysis between the COBE DMR 4 year, 
and ROSAT PSPC All-Sky Survey data are presented.
Statistical comparisons between microwave and X-ray maps 
can probe interesting astrophysical
environments and processes, such as the warm interstellar medium,
the Sunyaev-Zel'dovich effect in clusters of galaxies or gaseous 
group halos,
X-ray luminous radio sources and the Integrated Sachs-Wolfe
or the Rees-Sciama effect.
In order to test the diffuse, extragalactic X-ray background as probed by 
ROSAT, against the COBE DMR large-scale CMB structure, our analysis was
performed in most detail in a ROSAT selected region of the
sky (\mbox{$+\,40^\circ <$ b}, \mbox{$70^\circ <$ l $<
250^\circ$})
in an X-ray energy range with minimal Galactic structure and residual X-ray 
contamination, and the COBE low noise and least Galactic contribution 
channels.
Comparing to other regions of the sky and neighbouring 
energies and frequencies, 
we find indication for a positive Galactic correlation on large scales. 
This correlation 
is most prominent in the softest X-ray band and lowest microwave
channel, with a $>$ 95 \% confidence level detection against 
COBE noise and CMB cosmic variance, including the high quadrupole
value resulting from the power spectrum fit. 
The spectral dependences 
are consistent with Galactic thermal X-ray emission, 
and Galactic synchrotron radiation or free-free (Bremsstrahlung)
emission by the warm interstellar medium in the microwave regime.
Removing the quadrupole term on a sky map with a
Galactic cutout or related gradients in the selected
regions leaves no correlations above a 1-$\sigma$ level 
on smaller angular scales.
We conclude that there is no significant extragalactic correlation on scales 
for which the combined data are sensitive ($7^\circ - 40^\circ$) and 
that Galactic correlation is significant only on large angular scales, 
of the order of the quadrupole.
In the context of removing large angular scale gradients 
we give results on best fit X-ray dipoles from various
ROSAT data and discuss these with respect to the difficulty of finding 
a cosmological dipole.
The lowest correlation upper limits we can place are 
$\sim$ 15\% of COBE CMB fluctuations and $\sim$ 10\% 
of the ROSAT extragalactic XRB fluctuations. 
We discuss these results with respect to the possible correlation 
mechanisms. 


\end{abstract}
%________________________________________________________________
\section{Introduction}
  
Astrophysical radiation backgrounds are an important probe in 
cosmology. The instrument data used in this analysis, taken by the 
Cosmic Background Explorer's (COBE) Differential Microwave Radiometer
(DMR) (Smoot et al. 1990) and the R\"ontgen Satellite's (ROSAT)
(Tr\"umper et al. 1990) Position Sensitive Proportional Counter (PSPC) 
(Pfeffermann et al. 1986) All-Sky Survey (RASS) have contributed to
the improvement of our knowledge 
of the cosmic microwave (CMB) and the cosmic X-ray background
(XRB). The COBE DMR discovered for the first time structure of 
cosmic origin in the CMB (Smoot et al. 1992). The ROSAT PSPC 
data helped to resolve a large fraction (more than 75\%) of the 
emissivity of the XRB into point sources with redshifts out to 
$z \sim 3$ (Hasinger et al. 1993, Comastri et al. 1995).

As the two experiments probe very different energy regimes
($T_{CMB}$ = 2.7 K, $T_{1 keV} = 10^7$ K, Planck temperatures)
and the backgrounds originate at different redshifts
($z_{CMB} \sim 1000$, $z_{XRB} \sim$ 0--3) no strong correlations
can be expected between the two data sets. However physical processes
connecting the microwave and X-ray radiation and common emitting
sources do exist, leaving the possibility for a weak correlation.

Both data sets show the Galactic plane prominently and even at
high Galactic latitude Galactic emission is present in the maps
(Kogut et al. 1996). Correlation in this case can
mean spatial alignment of the emitting regions or common physical
processes. Galactic contributions to the microwave radiation in the DMR
frequencies are synchrotron radiation in the
Galactic magnetic field, dust radiation at $18$ K, and Bremsstrahlung from 
the warm interstellar medium of $8000$ K (Bennett et al. 1992). 
The diffuse Galactic X-ray
radiation consists of continuum and line emission from hot plasma of
1 to several million K. Structures are mainly constituted by the 
emission of nearby supernova remnants and superbubbles in the 
harder bands ($\sim$ 1 keV) 
but also from absorption through neutral gas associated with
H{\sc i} clouds in the softer bands ($\lsim$ 0.5 keV) (Snowden et al. 1995;
Egger et al. 1996).

An astrophysical environment directly connecting microwave and X-ray
radiation is a hot plasma with high column density along the CMB photon path, 
as was first noticed by Sunyaev \& Zel'dovich (1971) (SZ). 
In clusters of galaxies, electrons with
temperatures of $\sim 10^8$ K and densities of $\sim 3 \times 10^{-3}$
cm$^{-3}$ scatter photons of the CMB in the inverse Compton process
to higher energies leading to $y$-distortions of the spectrum and a
decrease in photon number in the Rayleigh-Jeans regime. The same
plasma, on the other hand, strongly emits at Bremsstrahlung energies
of 10 keV. Although the dependences on electron temperature $T_e$ and
electron density $n_e$ are different for the temperature decrement,
$\Delta T_{SZ}$, and the X-ray flux, $S_{X}$, according to the
approximate relations
\begin{displaymath}
\begin{array}{rcl}
\Delta T_{SZ} & \propto & n_e \, T_e \\
S_{X} & \propto & n_e^{2} \, T_e^{-\frac{1}{2}} \\
\end{array}
\end{displaymath}
a strong anticorrelation is present and has been observed in
individual clusters (e.g. Birkinshaw et al. 1991).

Recently Suto et al. (1996) suggested a similar 
effect for a possible plasma halo of the Local Group. As the assumed 
temperature ($\sim$ 1 keV) and density ($\sim 10^{-4}$ cm$^{-3}$) are 
lower, the expected effect is smaller than for galaxy clusters, 
but has a predicted angular structure. 
In the COBE DMR data this structure could not be found 
(Banday \& G\'orski 1996). A combination with the ROSAT X-ray
template, sensitive in this energy range, further improves the 
limits, also testing for a non-spherical structure of the halo.

Since the XRB does consist to a large extent of point sources, of
which the largest fraction are active galaxies (Hasinger et al. 1993),
a considerable number are also radio sources, e.g. radio-loud AGN. 
Laurent-Muehleisen et al. (1996), for example, find 2,127 sources common to
the RASS and the Green Bank 5 GHz radio catalog. 
Correlation analysis can thus test the fraction and angular scale 
of correlation at which these populations contribute to the data.

Also, recently Turok \& Crittenden (1996) proposed 
the Integrated Sachs-Wolfe (1967) (ISW) effect in a non-flat cosmological 
model as a possible source for a correlation between CMB and XRB data. 
This effect, for the non-linear evolution of a single gravitational
potential well, also called Rees-Sciama (1968) effect, introduces
anisotropies into the CMB via the change of the potential in time
$\dot\Phi$ along the photon path in the direction ${\mathbf n}$
between recombination $\tau_{rec}$ and reception $\tau_{0}$ time.
\begin{equation}
\frac{\delta T}{T} ({\mathbf n}) = 2 \int^{\tau_0}_{\tau_{rec}}
\dot\Phi(\tau,{\mathbf n}(\tau_0 - \tau)) \, d\tau.
\end{equation}
They calculated the linear evolutionary effect in detail for 
different $\Lambda$-models, with calculations for an open model
added by Kamionkowski (1996), and found the largest contribution at 
z $<$ 2. Thus, they suggest the distribution of sources in the
unresolved XRB to be a tracer of the potential for the redshift interval
of interest. If the redshift distribution and the biasing of the sources
is known, then a detected positive correlation can quantify $\Lambda$.
A summary of all these effects (table 2) and a discussion with respect to
the results of the correlation analysis is given in section 5.

A statistical comparison between observations of the CMB and the XRB 
was first carried out by Boughn \& Jahoda (1993) comparing the 19.2 GHz
survey with HEAO-1 A2 ($\sim$10 keV, $3^\circ$ resolution), 
and they found no significant correlation, based on Monte Carlo 
simulations for noise properties. Bennett et al. (1993) in cross-correlating 
the 1 year DMR data to HEAO-1 found no significant correlation 
for $|b| > 30^\circ$ and with the LMC masked. 
In the 4 year DMR data analysis by Banday et al. (1996) an 
expansion in orthogonal functions on a cut sky and a likelihood analysis 
for the coupling constant between the DMR and HEAO-1 data was
used in a simultaneous fit to the CMB power spectrum. 
Again, no significant correlation was found, when applying a specially 
designed Galactic cut based on correlations obtained from 
the DIRBE 140 $\mu$m map and masking of the LMC.
Using the ROSAT PSPC extends previous work to softer X-ray energies, 
higher angular resolution and better sensitivity.

Additional interest in this analysis arises from the detection 
of a spatially extended X-ray source around clusters of galaxies 
found in a correlation analysis by So\l tan et al. (1996a) between 
Abell clusters and the ROSAT diffuse XRB. Correlating to the COBE DMR 
can constrain a gas halo model for the extended component.

In section 2 we introduce the COBE and ROSAT data sets
respectively used for
this analysis and explain how they were prepared. In section 3 
the correlation method is described, including our error estimation. 
Section 4 contains the results on various angular scales 
and their dependences with energy or frequency, and in section 5 we 
discuss the results in the context of theoretical expectations of 
possible correlation mechanisms. 

\section{The Data}

\subsection{COBE DMR}

The COBE DMR measures the sky differentially in 3 frequency channels
(31.5, 53 \& 90 GHz). The maps are binned in $2.6^\circ \times
2.6^\circ$ pixels, which are considerably smaller than the 
beam width of 7$^\circ$ (FWHM). 
The CMB signal has an amplitude of $35 \pm 2 \mu$K and 
the COBE DMR mean sensitivity is 26 $\mu$K per resolution element, 
corresponding to a signal to noise ratio of $\sim$ 0.5 on pixel level.
The noise level varies from channel to channel
(31A: 248, 31B: 316, 53A: 86, 53B: 101, 90A: 146, 90B:
116; in $\mu$K per pixel).
The results of the COBE 4 year data
analysis by the COBE team are summarized and referenced in 
Bennett et al. (1996).
We used various maps of the {\it COBE} DMR data set. The final
analysis uses the 4 year data, but consistency tests with the 1 year and 
2 year data were performed. From the channels A and B, which are
differentially measuring the signal, we
constructed inverse noise weighted (A+B)/2 and frequency combined sum
maps to minimize the noise.
\begin{equation}
\Delta T_{ij} = \frac{1}{W_{ij}} (w_{i} \, \Delta T_{i} + w_{j}
\, \Delta T_{j})
\end{equation}
with $w_\star = \sum_\star 1/\sigma_\star^2$ and $1/W_{ij} = 1/(w_{i}+w_{j})$, 
where $w_\star$ is evaluated on the custom cut sky (Banday et al. 1997), 
considering the noise level and exposure.

All maps used were converted from antenna to Planck
temperatures. The standard frequency combination for 
our analysis is the 53+90 GHz map, with low noise and little Galactic 
contribution (Kogut et al. 1996). The individual frequency maps and 
the linearly combined galaxy reduced maps (cmb \& smb) were used for 
comparison.

\subsection{ROSAT PSPC All-Sky Survey}

The ROSAT satellite covers a large energy range (0.1 -- 2 keV) 
in the soft X-rays. The harder part is divided into 4 bands with 
maximum responses at the following energies 
(R4: 0.7 keV, R5: 0.8 keV, R6: 1.1 keV, R7: 1.5 keV). 
All bands, particularly the neighbouring ones, have considerable overlap 
with each other of up to 50\%, due to the limited spectral resolution of
the proportional counters.

The RASS intensity (I) and noise ($\sigma$) maps were constructed as 
\begin{equation}
I = \frac{Ct - B}{Ex} 
\end{equation}
\begin{equation}
\sigma = \frac{\sqrt{Ct}}{Ex} 
\end{equation}
The abbreviations denote count number of received photons
($Ct$), modeled contamination ($B$) and
exposure ($Ex$). The X-ray contamination in ROSAT consists, with varying
contribution in the different energy bands, of solar scattered
X-rays, ``short- and long-term enhancements'' and particle
background. The noise maps are calculated according to Poisson 
statistics due to photon number limitation.
The energy band R6 (0.73 -- 1.56\,keV) is regarded as the best probe for the 
diffuse cosmological XRB, because the systematic uncertainty 
induced by foregrounds, such as 
non-cosmic photons, and contamination by charged particles is
minimized. We concentrate our analysis and results to this band, 
but investigate systematic effects by comparison with the other hard
bands. Particularly the R5 band is also low in non-cosmic photons and 
in contamination by charged particles, but which contains, compared to the R6
band, increased Galactic foreground and can thus hint at a discrimination
between the Galactic and the extragalactic signal. 
In spite of the careful corrections 
for exposure and elimination of non-cosmic backgrounds
(Snowden et al. 1995) the final count rate distribution 
is not completely free from residual contamination.
We therefore tested extensively for correlations induced by exposure or 
contamination corrections and found no negative effects on the 
results. Unlike the RASS data described in Snowden et
al. (1995), the data set used here has been constructed on a ``photon
by photon'' basis, i.e. the intrinsic resolution is that of the
detector ($\sim 1'$). Thus, the sensitivity has been improved by
avoiding ``crosstalk'' from bright sources.

The maps used in our work were binned into 
0.7$^\circ$ x 0.7$^\circ$, with point sources included, in order to compare to 
the complete integrated flux. In a second step they were rebinned to
COBE DMR pixel size with varying exclusion thresholds for bright sources.
The mean intensity of the XRB 
in the R6 is $\sim$
1.9 cts s$^{-1}$ per pixel
with a fluctuation level of $\sim$ 
0.17 cts s$^{-1}$ per pixel
and a signal to noise ratio of $\sim$ 2 at COBE DMR pixelization.

Even the high energy R6 band is, in large regions of the sky, 
dominated by Galactic emission. A field almost free from Galactic
structures, which is sufficiently large, was
chosen (\mbox{$+\,40^\circ <$ b}, \mbox{$70^\circ <$ l $<
250^\circ$}), hereafter called the selected NGP field.
This field is the largest simply connected patch of the sky probing 
primarily the XRB. Properties of the XRB in this field have been 
studied in a series of papers (So\l tan et al. 1996a; Miyaji et
al. 1996; So\l tan et al. 1996b).

\section{Correlation Method}

\subsection{Correlation Function}

To minimize the Galactic contribution, 
a patch of the sky which has the highest 
sensitivity to the diffuse, extragalactic XRB was chosen. Due to the size 
($\sim 8\%$ of the sky) and the peculiar geometry of the patch, 
a local statistical measure for similarities in structure, 
the 2-point correlation function, is 
preferred over global measures such as correlated power spectrum
components in e.g. a spherical harmonic expansion. The form of the 
correlation function used in this analysis is the Pearson product
moment correlation coefficient $C(\alpha) =$
\begin{equation}
\frac{<X_i T_j>_\alpha \, - \, <X_i>_\alpha  
\, <T_j>_\alpha}{\sqrt{<X_i^2>_\alpha \, - \, <X_i>_\alpha^2} \,\,\,  
\sqrt{<T_j^2>_\alpha \, - \, <T_j>_\alpha^2}} 
\end{equation}
in an unweighted scheme. Inverse noise variance weighting has also been 
used and was found to give unchanged results. 
The correlation coefficient varies between 1 and -1 for correlation
and anticorrelation respectively with 0 indicating no correlation. 
For pixel sizes comparable to the resolution limit, the statistical 
uncertainty of the measure is $( 1 - C^2(\alpha) ) \, \, 
\sqrt{N_{\{ij\}}-1}$. 
The average is taken over 
all pixel pairs $\{ij\}$ with separation $\alpha$ in the patch. The 
subscript $\alpha$ denotes that all the terms were evaluated
separately for each angle bin. This ensures a correctly weighted 
normalization even in cases when, due to limited area and boundary
effects, the zero-lag field properties are not a fair ensemble average 
for all angle bins any longer. 
Note that through our choice of the cross-correlation function, 
the bins are completely statistically independent, and hence 
any apparent correlation between the bins is due to the structures 
in the maps. 

To determine the uncertainties, which we assume to be dominated by 
the DMR noise (section 2) and the cosmic variance of the CMB
structure, different techniques were applied. We
applied a simple method, 
which introduces little prejudice (just assumes rotational
invariance of the data), to correlate to random samples drawn from
the maps by rotation. Since the ROSAT maps are known to contain a strong 
Galactic, not rotationally invariant contribution 
on major parts of the sky, in contrast to the
COBE maps, which have been investigated and found to be primarily
consisting of CMB structures (Kogut et al. 1996), 
and since the COBE data predominantly introduce the errors, 
those were rotated around the NPG, 
and mirror image rotated around the SGP in $10^\circ$ steps to
produce 35 and 36 random samples each. The error estimates induced 
by this method agree well with our second method using simulations of 
DMR maps. The CMB 
structure was taken to be a random Gaussian field on the sky 
with a power-law 
($Q_{rms-PS}$ = 15.3 $\mu$K, n = 1.2) power spectrum (G\'orski et
al. 1996) convolved with the 
DMR filter function (Kneissl \& Smoot 1993). The modes used to
construct the map cover a range in multipole index $\ell =$ 2--25. 
Our results were compared against $\sim$ 1000 simulations.
The DMR noise is given as Gaussian pixel noise distributed according 
to the coverage.

\subsection{Spherical Harmonic Fit}

Although we use a correlation function analysis and not a power
spectrum correlation analysis, it is important to study 
how the correlation results are influenced by particularly the low 
order multipoles. For this we expand the sky maps 
\begin{equation}
X( \vartheta , \varphi ) = \sum_{\ell=0}^{\infty}
\sum_{m=-\ell}^{+\ell} b_{\ell m} Y_{\ell m}( \vartheta , \varphi ) 
\end{equation}
into real valued spherical harmonics 
\begin{displaymath}
\displaystyle{\begin{array}{lc}
Y_{\ell m} ( \vartheta , \varphi ) = & \\
 & \\
\sqrt{\frac{2\ell + 1}{2\pi}} \, 
\sqrt{\frac{(\ell - |m|)!}{(\ell + |m|)!}} \, P_\ell^{|m|}(\cos\vartheta )%}
\left\{ 
\begin{array}{lr}
\sin |m| \varphi, & m < 0 \\
\frac{1}{\sqrt{2}}, & = 0 \\
\cos |m| \varphi, & > 0 
\end{array} \right. & \\
\end{array}}
\end{displaymath}

\begin{equation}
\displaystyle{P_\ell^m (x) \, = \, 
{(-1)}^m \, (1 - x^2 )^{-\frac{m}{2}} 
\, \frac{d^m}{dx^m} P_\ell (x)}, \qquad m > 0
\end{equation}
\begin{displaymath}
\displaystyle{P_\ell (x) \, = \, \frac{1}{2^\ell \, \ell !} \, 
\frac{d^\ell}{dx^\ell} (x^2 - 1)^\ell}
\end{displaymath}
in a simultaneous fit up to multipole order $\ell_{max}$, and subtract 
these multipoles from the maps. A well-known problem (e.g. Bunn et al.
1994) arises in this procedure. The orthogonality relation for the 
spherical harmonics does not hold in the case of incomplete sky
coverage: 
\begin{equation}
\int_{R} \, Y_{\ell m}( \Omega ) \, Y_{\ell'
m'}( \Omega ) \, d \Omega = W_{\ell \ell' m m'}, 
\end{equation}
where $W_{\ell \ell' m m'} \ne \delta_{\ell \ell'} \delta_{m m'}$ in
general, if $R \subset S_{1}$.
In our case this leads to the fact that we subtract a function from 
the sky which can be expressed as a sum of different multipoles. 
Nevertheless the order of angular scale of variation for this function 
is similar to the dominant multipole probed. Only in close comparison 
with theoretical power spectrum estimates is an exact determination 
of the individual multipole terms of interest. For this we
investigated the amount of ``cross-talk'' between the modes by 
varying $\ell_{max}$ to the stability limits of the fit and 
studying the noticeable changes of the subtracted multipole modes, 
since the limitation to a range in $\ell$-space is the major problem 
for the fit technique. The changes to the results of our correlation 
analysis and dipole determination turn out to be insignificant at 
the present sensitivity level. In the case of fitting a dipole 
to a field the resulting parameters describe the direction and 
amplitude of a local gradient and can only be
compared for consistency with a whole sky dipole. In the case of the 
correlation analysis the errors induced by the effect are
statistically taken into account by subjecting the simulated data 
to the same subtraction procedure. We find good agreement between 
the best fit multipoles and expected amplitudes from the correlation results 
on the region, or a subset of the region for which the fit is
constrained. This, combined with the fact that 
the correlation function seems to be a fairly unbiased estimator 
for the amplitude of multipoles (Bunn et al. 1994), 
at least in the case of a power-law model for the CMB fluctuations, gives 
us further confidence in the validity of the method used and 
we did not see the need to apply more sophisticated methods
such as, e.g. constructing orthogonalized functions on parts 
of the sky (G\'orski 1994) or utilizing model assumptions in 
Wiener filtering techniques (Zaroubi et al. 1995), in the present 
state of the work. 

\subsection{Quantifying Results}

For quantifying limits on the strength of the correlation ($\beta$), 
we use a method which evaluates a 
likelihood distribution for $\beta$, see e.g. Bennett et al. (1993).
Assuming that the signal in the ROSAT data (X) and the
COBE data (T) can be written in pixel
space, leaving out the index $i$, as
\begin{equation}
\begin{array}{rcl}
T & = & T_{CMB} + T_{noise} + \beta_{T|X} X \\
 & & \\
X & = & X_{XRB} + X_{noise} + \beta_{X|T} T \\
\end{array}
\end{equation}
where $\beta_{T|X}$ and $\beta_{X|T}$ are the regression coefficients
which 
couple the two maps by regarding one as template for the other. 
Clearly the COBE noise and CMB cosmic variance dominating the
temperature distribution are the main source of confusion for the
cross-correlation. The Poisson noise in ROSAT by 
photon number limitation is apparently a 
small confusion term. Shot noise is hard to distinguish from the XRB 
since we are interested in a possible correlation of ROSAT sources. 
From the results of testing with different source exclusion
thresholds, we deduce that shot noise is small after 
exclusion of a few very strong sources. A probably considerable 
confusion term for the cross-correlation is chance alignment of the
structure in the diffuse XRB, which we denoted by $X_{XRB}$ 
(here meaning the uncorrelated part). 
Due to the complexity of the diffuse XRB, which consists of various
typs of sources, 
more so than the CMB, no established model for the fluctuations exists. 
There is interesting work (e.g. Lahav et al. 1996)
to model the extragalactic XRB fluctuations, 
however, these models have not yet been compared to data and 
hence are still somewhat uncertain.
We believe that overall, the COBE noise and 
CMB cosmic variance correlated with the real structure in the ROSAT
maps are the dominant confusion terms, and for now, we 
have not attempted to model the XRB. 
Forming the correlation functions then yields the relations 
\begin{equation}
\begin{array}{rcl}
<XT> & = & <XT_{CMB}> + <XT_{noise}> + \\
 & & \beta_{T|X} <XX> \\
 & & \\
<TX> & = & <TX_{XRB}> + <TX_{noise}> + \\ 
 & & \beta_{X|T} <TT> \\
\end{array}
\end{equation}
where $<XT_{CMB}>$, $<XT_{noise}>$, $<TX_{XRB}>$ \& $<TX_{noise}>$
are assumed to be zero in the sense of a statistical average. 
This leads to the approximate relations
\begin{equation}
\begin{array}{rcl}
\beta_{T|X} & \sim & <XT>_{0^{\circ}} / <XX>_{0^{\circ}} \\
 & & \\
\beta_{X|T} & \sim & <XT>_{0^{\circ}} / <TT>_{0^{\circ}} \,\, .\\
\end{array}
\end{equation}
For one particular realization this is only true within some error, 
which may be determined through Monte-Carlo simulations
assuming models for $T_{CMB}$, $T_{noise}$, $X_{XRB}$ and $X_{noise}$.
We also take account of statistical biasing, which turned out to be small 
in comparison to the random errors. 
We assume now that $T_{CMB}$ and $T_{noise}$ are the dominant sources 
of error compared to Poisson noise and random structure in the XRB, and
approximate $< T (X_{XRB}+X_{noise}) >$ by $< (T_{CMB}+T_{noise}) X >$. 
Our method of determining $\beta_{T|X}$ and the corresponding 
uncertainty is to minimize 
\begin{eqnarray}
\chi^2 \, = \sum_{kl} \,\, (<TX> - \beta_{T|X} <XX>)^{T}_{k} \,\,\,\,
M_{kl}^{-1} \nonumber\\
 (<TX> - \beta_{T|X} <XX>)_l
\end{eqnarray}
with $M_{kl} \, = \, < X (T_{CMB}+T_{noise}) >_{kl}$ which can be determined
from a distribution of realizations of the model. 
Assuming Gaussian
errors, the probability distribution for $\beta$ can be drawn from the
$\chi^2$-distribution as 
\begin{equation}
P(\beta) \, d\beta \,\, \propto \,\, e^{{- \frac{1}{2} \chi^2}} \, d\beta.
\end{equation}

\section{Results on Various Scales}

Since the aim of this analysis is to compare a CMB measurement to the 
diffuse XRB, some individual very strong X-ray point sources, which
could influence even a statistical comparison, are removed from the maps. 
Furthermore, the influence of different point source exclusion thresholds 
on the results were studied and found to be mostly insignificant.
In a few cases, e.g. MK 421, a BL Lac object at 
z $\sim$ 0.031, the removal of the source lead to a
decrease of the correlation signal on the DMR beam scale. 
Comparison with known strong radio point sources showed no significant
contribution to DMR (Kogut et al. 1994), so chance alignment seems 
a likely cause. Nevertheless, systematic comparison between COBE data and
candidate radio-loud X-ray point sources seems a worthwhile check. 
The best fit DMR residual
dipole, which would introduce substantial correlation of no physical 
significance, had been removed, in addition to the standard removed dipole.

\subsection{Selected NGP Field}

\begin{figure}[htb]
\psfig{file=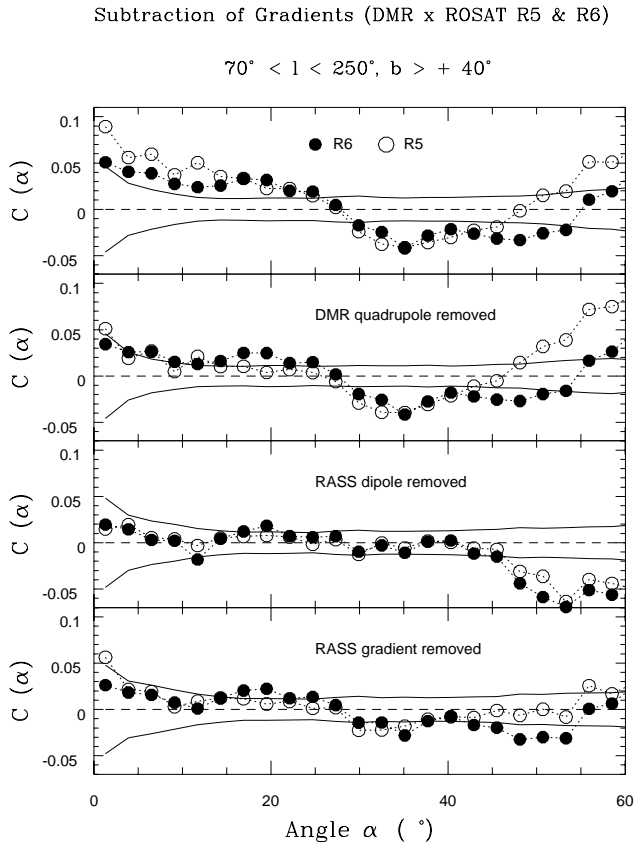,width=\hsize,clip=}
\caption{Cross-correlation function between the COBE DMR 53+90 GHz
and the ROSAT bands R6 (black) and R5 (white) in the selected
NGP field 
(\mbox{$+\,40^\circ <$ b}, \mbox{$70^\circ <$ l $< 250^\circ$}).
The effect of lowering the zero-lag amplitude 
by subtracting different multipoles related to a gradient 
on the field is shown (lowest panel). 
Subtracted are a best fit quadrupole
from the Galactic custom cut DMR map, a best fit dipole which has Galactic
signature from the Galactic cut ROSAT map ($|$b$| >$ 20$^\circ$)
and a gradient as a dipole
fitted on the field. The two latter both have Galactic signature 
(table 1).
The 1-$\sigma$ error bands are taken from DMR noise + 
CMB simulations correlated to the ROSAT energy band R6.}
\label{fig:}
\end{figure}

Correlating the raw data (without excluding source contaminated pixels
or subtracting structure) yields a marginally significant positive 
correlation on large angular scales (figure 1), 
which appeared to be independent of the following different 
procedures that had been applied. 
Different source exclusion thresholds in the ROSAT maps ranging 
from 0.3--1 cts s$^{-1}$  were compared. 
After excluding the strong source MKN 421 with 5.3 cts s$^{-1}$ in the
R6 energy band, the results were only marginally affected by different 
thresholds. Different sampling tests were undertaken, also showing
stability of the result against small scale features such as point
sources and noise. The maps were smoothed on various angular scales
including smoothing of the ROSAT maps with the actual DMR beam
(Kneissl \& Smoot 1993), and Gaussian
smoothing of both maps out to $20^\circ$, with the effect of smoothing 
the correlation function, but not significantly changing the
correlated signal. 

The energy dependence in X-rays is found to increase from 
hard to soft energies. The frequency dependence in microwaves is
somewhat unclear. There is a clear signal in both the 
53 and 90 GHz channels and no signal at 31.5 GHz.

To determine the angular scale of the correlated signal, gradients were 
removed from the field. This was done in fitting dipoles onto 
the field in both maps and subtracting them. As a result the signal 
is reduced below the 1-$\sigma$ level. 
The multipoles on the sky dominating these gradients 
turn out to be of low order (figure 1). 

The gradient in the ROSAT selected NGP field has similar 
orientation as the whole map dipole (table 1), the positive pole lying near 
the Galactic center, and increasing in relative amplitude from hard to 
soft energies. From this energy dependence we derive spectral
properties of the emitter that are in agreement with a $2 \times
10^{6}$K equilibrium plasma, typical for Galactic emission. 
The energy dependence, however, is not compatible with the X-ray
spectrum of the extragalactic XRB (Hasinger et
al. 1993). 

The gradient in the COBE 
field is dominated by a quadrupole fitted to the COBE cut sky, a
combination of the cosmic and the Galactic quadrupole. 
In the field, the cosmic quadrupole seems to dominate, which would not
be inconsistent with the COBE frequency dependence of the correlated signal.
A signal constant with frequency would be expected, but the 31.5 GHz
channel could be confused by the increased noise level and Galactic
contribution. 

Removing the gradients no significant correlation is left, and we can
set upper limits of 4.5 $\mu$K and 0.02 cts s$^{-1}$ per pixel
(95 \% CL) in the R6 band on a correlation between the 
CMB and the extragalactic XRB on scales of $7^\circ - 40^\circ$.

\subsection{Cut Sky}

\begin{figure}[htb]
\psfig{file=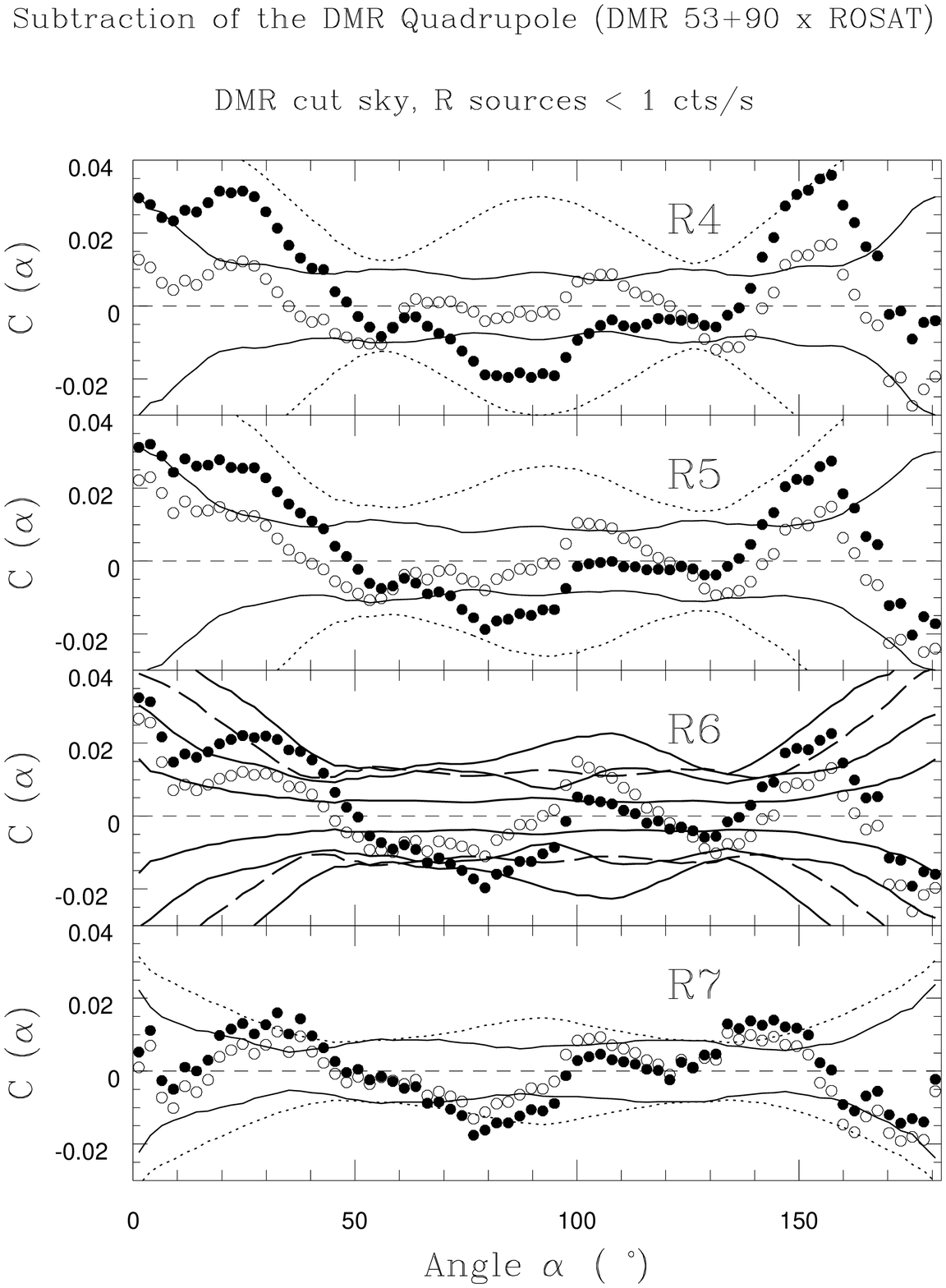,width=\hsize,clip=}
\caption{Cross-correlation between DMR 53+90 GHz and
different ROSAT energy bands including (black) and excluding
(white) the DMR quadrupole. The DMR and ROSAT data are both taken 
from the same region, i.e. the DMR Galactic cut sky. 
Apart from the R6 band the narrow (solid line) and  
wide (dotted line) 1-$\sigma$ error bands are taken from DMR noise 
and CMB power spectrum simulations excluding and including
the variance from the $Q_{rms-PS}$ = 15.3 $\mu$K respectively. 
In the R6 band panel we demonstrate the 1-$\sigma$ error bands 
from (with increasing amplitude at zero-lag): DMR noise only, noise + 
CMB ($\ell > 2$), noise + CMB ($\ell > 2$) + $Q_{rms}$ (= 6 $\mu$K) 
(dashed line) and noise + CMB ($\ell > 1$). 
There is no significant correlation, but a trend of
increasing positive quadrupole correlation towards softer X-ray energy
bands with about 68 \% CL at R4 against the cosmic variance of 
$Q_{rms-PS}$ = 15.3 $\mu$K. The small excess in correlation on scales 
out to $10^{\circ}$ most prominent in DMR x R6 has been found to be 
partly due to the LMC.}
\end{figure}

\begin{figure}[htb]
\psfig{file=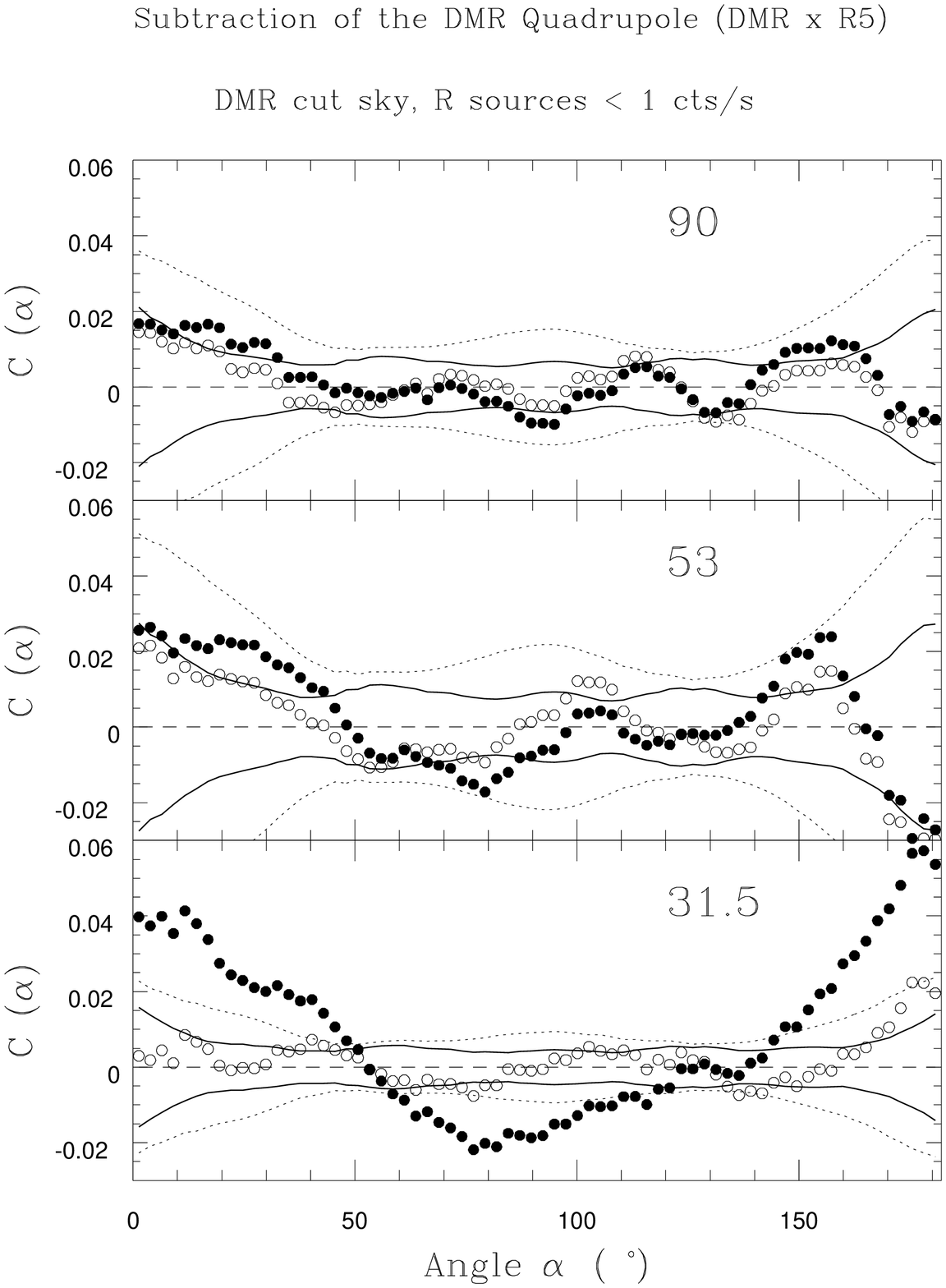,width=\hsize,clip=}
\caption{Cross-correlation between ROSAT R5 and the individual DMR
frequency channels (GHz) including (black) and excluding (white) the DMR
quadrupoles. The DMR and ROSAT data are both taken 
from the same region, i.e. the DMR Galactic cut sky. 
The 1-$\sigma$ error bands are taken from DMR noise 
and CMB simulations including and excluding $Q_{rms-PS}$ 
respectively. There is a 
trend of increasing quadrupole correlation towards lower DMR 
frequencies with a significant ($>$ 95 \% CL) detection at 31.5 GHz.} 
\end{figure}

For completeness we present the correlation function on a large fraction
of the sky, the DMR Galactic cut sky, although these results are not
relevant for a comparison with the extragalactic XRB. 
We use the DMR cut sky for both data sets, 
since the most interesting question in this case is that of a
correlation to Galactic features in the COBE DMR, which should be 
largely excluded by the custom cut (Banday et al. 1997), which is a 
straight 
$|b| > 20^\circ$ cut, with additional cutouts for Galactic structures
found in correlation to the 140 $\mu$m COBE DIRBE map (``flares of
obscuration'' in Scorpius, Ophiuchus, Taurus and Orion).
We additionally experimented with 
$|b| > 20^\circ$ and $30^\circ$, and found, not 
surprisingly, a slight, not very significant tendency towards positive 
correlation with smaller cut angle. To minimize confusion, we subtract 
before the correlation in the cut sky case, in addition to the 
best fit DMR dipole, also the best fit ROSAT dipole. 

The corresponding ROSAT sky is largely dominated by strong
Galactic features such as, e.g. the Loop I with the prominent North Polar
Spur, or the Eridanus enhancement, which are claimed not to be found
in the DMR maps (Kogut et al. 1996) and therefore are sources
of confusion.
The limits we get confirm this view. The limits 
on the contribution of ROSAT to COBE are stronger 
compared to the ones derived on the
selected NGP field, but this has to be taken with caution, 
since ROSAT is by a factor of $\sim 2$ less constrained, 
due to increased uncorrelated (Galactic) structure in ROSAT. 

In figure 2 we find the DMR quadrupole to be positively correlated to the 
ROSAT R4-6 templates. Here the outer 1-$\sigma$ error bands demonstrate 
the inclusion of the $Q_{rms-PS} = 15.3 \, \mu$K quadrupole. 
We stress here the fact that the huge cosmic variance of this term 
($\sim 68\%$) introduces a large error, which is problematic, 
as the actually measured, uncorrected 
DMR $Q_{rms}$ is considerably smaller (15.6$\pm$5.4$\mu$K [31.5
GHz], 4.4$\pm$3.3$\mu$K [53 GHz], 0.0$\pm$3.0$\mu$K [90 GHz]),
because of an apparently anti-correlated alignment of the CMB
quadrupole ($10.7 \pm 3.6($random$) \pm 7.1($systematic) $\mu$K) 
with the Galactic quadrupole (Kogut et al. 1996).
For the energy dependence of the effect, we find agreement with a 
thermal $2 \times 10^{6}$K spectrum and disagreement with an 
extragalactic spectrum (a power-law with a photon index $\Gamma\approx 
-2$).

Comparing the ROSAT R5 band to the different 
DMR frequencies (figure 3) we find a significant
positive correlation ($> 95$ \% CL)
at 31.5 GHz due to a quadrupole term
($Q^{COBE}_{rms} = 10.3 \pm 4 \, \mu$K, 
$Q^{ROSAT}_{rms} = 0.38 \pm 0.15$ cts s$^{-1}$ per pixel).
At the other frequencies a correlation due to the quadrupole is also
positive, but not significant compared with the full CMB spectrum
including the quadrupole. Investigating the spectral dependence 
we find the spectral index $\beta, T_{antenna} \propto \nu^{\beta}$, 
within 1 $\sigma$ to be consistent with $\beta_{ff} \sim -2.1$ and 
$\beta_{synchrotron} \sim -2.8$. 

When subtracting the best fit quadrupole, there is in no case any 
significant correlation left in the data. The limits are in the 
case of DMR 53+90 x ROSAT R6 on a $30^{\circ}$ cut sky 
$2.3 \, \mu$K and 
0.037 cts s$^{-1}$ per pixel (95 \% CL).

From the systematics of the quadrupole correlation, we conclude 
that a correlation between Galactic quadrupoles in ROSAT and COBE 
seems likely, but which however, apart from the COBE DMR 31.5 GHz channels, 
can not be shown with sufficient significance against the cosmic
variance of a quadrupole consistent with the CMB spectrum. 
We see this result in agreement with the existence 
of a Galactic DMR quadrupole (Bennett et al. 1992; Kogut et al. 1996). 

\subsection{Dipole}

\begin{table*}[htbp]
\begin{center}
\footnotesize
\begin{tabular}{|c|c c c|c c c|c c c|}
\hline
NGP & & & & & & & & & \\
70--250 & dof & $\chi^2$/dof & b$_{00}$ & $\tilde b_{1-1}$ &
$\tilde b_{10}$ & $\tilde b_{11}$ & $\ell^{II}$ & b$^{II}$ & D \\
src $<$ 0.45 & & & & & & & & & \\
\hline
R4 & 548 & 5.3 & 217$\pm$4 & 1.1$\pm$0.3 & 0.6$\pm$1.2 & 27.7$\pm$0.8
& 2.3 & 1.9 & 7.8 \\
R5 & 548 & 4.4 & 227$\pm$4 & -0.6$\pm$0.3 & 5.3$\pm$1.1 & $14.2\pm$0.6
& 357.6 & 27.8 & 4.3 \\
R6 & 548 & 5.4 & 272$\pm$4 & -0.5$\pm$0.3 & 3.3$\pm$1.0 & 5.3$\pm$0.5
& 354.7 & 40.8 & 1.8 \\
R7 & 548 & 3.4 & 146$\pm$4 & -0.9$\pm$0.5 & 6.4$\pm$1.7 & 1.0$\pm$0.8
& 316.4 & 81.5 & 1.8 \\
\hline
SGP & & & & & & & & & \\
70--250 & & & & & & & & & \\
src $<$ 0.45 & & & & & & & & & \\
\hline
R4 & 548 & 10.8 & 221$\pm$5 & -6.3$\pm$0.4 & 12.8$\pm$1.3 & 2.9$\pm$0.7 &
294.6 & 69.0 & 4.1 \\
R5 & 548 & 8.4 & 249$\pm$5 & -3.1$\pm$0.3 & 9.4$\pm$1.2 & -0.8$\pm$0.6 &
255.3 & 76.5 & 2.8 \\
R6 & 548 & 6.1 & 305$\pm$5 & -2.5$\pm$0.3 & 6.6$\pm$1.1 & 2.8$\pm$0.5 &
318.1 & 68.3 & 2.1 \\
R7 & 548 & 3.0 & 208$\pm$5 & -2.7$\pm$0.4 & 13.0$\pm$1.5 & 7.1$\pm$0.7 &
339.0 & 67.5 & 4.2 \\
\hline
N+SGP & & & & & & & & & \\
70--250 & & & & & & & & & \\
src $<$ 0.45 & & & & & & & & & \\
\hline
R4 & 1100 & 10.2 & 200$\pm$0.6 & -1.3$\pm$0.2 & 3.0$\pm$0.1 &
17.3$\pm$0.4 & 355.7 & 14.0 & 5.0 \\
R5 & 1100 & 7.2 & 231$\pm$0.6 & -1.4$\pm$0.2 & 2.2$\pm$0.1 &
7.6$\pm$0.4 & 349.4 & 21.8 & 2.3 \\
R6 & 1100 & 5.8 & 281$\pm$0.6 & -1.2$\pm$0.2 & 1.0$\pm$0.1 &
4.1$\pm$0.3 & 343.5 & 18.2 & 1.2 \\
R7 & 1100 & 3.3 & 162$\pm$0.6 & -1.6$\pm$0.3 & -0.4$\pm$0.1 &
3.1$\pm$0.5 & 332.6 & -8.7 & 1.0 \\
\hline
N+SGP & & & & & & & & & \\
70--250 & & & & & & & & & \\
src $<$ 1 & & & & & & & & & \\
\hline
R4 & 1100 & 10.4 & 201$\pm$0.6 & -1.4$\pm$0.2 & 3.1$\pm$0.1 &
17.5$\pm$0.4 & 355.4 & 14.0 & 5.0 \\
R5 & 1100 & 8.4 & 232$\pm$0.6 & -1.5$\pm$0.2 & 2.2$\pm$0.1 &
8.0$\pm$0.4 & 349.4 & 21.0 & 2.4 \\
R6 & 1100 & 7.3 & 282$\pm$0.6 & -1.1$\pm$0.2 & 1.0$\pm$0.1 &
3.9$\pm$0.3 & 344.9 & 19.6 & 1.2 \\
R7 & 1100 & 3.5 & 163$\pm$0.5 & -1.7$\pm$0.3 & -0.4$\pm$0.1 &
3.3$\pm$0.5 & 333.4 & -8.2 & 1.1 \\
\hline
cut sky & & & & & & & & & \\
($|b|>$20) & & & & & & & & & \\
src $<$ 1 & & & & & & & & & \\
\hline
R4 & 4012 & 36.8 & 231.0$\pm$0.2 & -4.3$\pm$0.1 & 5.9$\pm$0.1 &
30.0$\pm$0.1 & 351.8 & 15.5 & 8.7 \\
R5 & 4012 & 54.3 & 275.7$\pm$0.2 & -5.3$\pm$0.1 & 4.2$\pm$0.1 &
26.4$\pm$0.1 & 348.7 & 12.4 & 7.7 \\
R6 & 4012 & 34.8 & 315.9$\pm$0.2 & -4.2$\pm$0.1 & 2.5$\pm$0.1 &
16.9$\pm$0.1 & 346.1 & 11.6 & 5.0 \\
R7 & 4012 & 7.1 & 169.3$\pm$0.2 & -3.5$\pm$0.1 & 0.6$\pm$0.1 &
6.0$\pm$0.1 & 329.4 & 7.2 & 2.0 \\
\hline
\end{tabular}
\normalsize
\caption{Dipole fits to various subsets of the ROSAT PSPC All-Sky
Survey data. 
The fields NGP (b $>+40^\circ$) and SGP (b $<-40^\circ$) are in the
longitude ranges indicated. The source exclusion thresholds (src) are in cts
s$^{-1}$. b$_{\ell m}$ are the coefficients of the real valued spherical
harmonics, whereas $\tilde b_{1m}$ is $100 \times b_{1m}$ / $b_{00}$, 
the dipole coefficients in percentage of the monopole term. 
($\ell^{II}$, b$^{II}$) gives the best fit position for the positive
pole in Galactic coordinates and D$^2 = 1/(4\pi )
\sum_{m=-1}^{1} \tilde b_{1m}^2$ is the amplitude. Although we show
the (small) fit errors on the $\tilde b_{1m}$ resulting from ROSAT
noise only, we do not translate them into the coordinate values 
as we see the results dominated by systematic errors. 
For details see text.} 
\end{center}
\end{table*}

The COBE dipole is the well known Doppler dipole 
due to the sun's motion with respect to the CMB rest frame. Any 
other residual dipole is an inseparable combination of imperfect
Doppler dipole subtraction, Galactic dipole and CMB dipole. Thus a 
best fit dipole has to be removed from the maps, because the induced 
correlation has no physical interpretation and thus confuses the 
results. Still the best fit dipoles in ROSAT can be determined and 
simply compared to amplitude and orientation of e.g. the DMR Doppler
dipole, an expected Galactic dipole, etc.

The dominant part of the correlated signal on the selected NGP field can be
explained by a Galactic gradient in ROSAT. Comparing orientation and 
amplitude of this gradient expressed as a dipole on the sky, to dipoles
fitted to other regions (table 1), we find it to lie in the general
direction of the Galactic center and North Polar Spur region 
and to have an amplitude of the dipole component towards 
the Galactic center ($\tilde b_{11}$) which significantly increases 
towards softer X-rays, consistent with a Galactic energy spectrum. 
An increase in Galactic latitude towards harder X-rays in the selected
NPG field, although also preferred by fitting in the SGP field, 
has to be attributed to local Galactic phenomena, 
since it is not supported by a more global, joint
fit to both fields combined. The $\chi^2/$degrees of freedom (dof)
indicates that the gradients / dipoles are clearly not the 
dominant structures, but 
they are fairly well defined in terms of the formal fit errors and 
insensitive to point source contributions (compare panels 3 and
4 of table 1). For the determination of an extragalactic global dipole, 
systematic errors induced by Galactic features appear to be the 
overwhelming source of confusion. The dipole fits presented here can be 
compared to work by Freyberg et al. (1996), who investigated the
ROSAT data with regards to a possible Galactic X-ray halo.

\section{Discussion}

\subsection{Dipole}

\begin{table*}[htbp]
\begin{center}
\begin{tabular}{|l|c|c|c|c|l|}
\hline
 & & & & & \\
 {\it EFFECT /} & {\it SIGN} & {\it ANGULAR} & {\it
FREQU. DEP.} & {\it ENERGY DEP.} &
{\it AUTHOR} \\
 {\it SOURCE} & & {\it SCALE} & microwave & X-ray & \\
 & & & & & \\
\hline
\hline
 & & & & & \\
  {\bf Galaxy} & + & large & $\beta_{synch,ff,dust} \approx $ & thermal & \\
 geometrical ? & ($-$) X-abs. & & $\{ -2.8,-2.1,1.5 \}$ & 0.3 keV & \\
 & & & & & \\
\hline
 & & & & & \\
 {\bf SZ thermal} & $-$ & ($<$ 10' clust./ & y - distortion & thermal & \\
 ({\bf clusters} / super- )& (Rayleigh--Jeans) & $\sils 5^\circ$ c-corr.) & & 10 keV & \\
 & & & & & \\
\hline
 & & & & & \\
 {\bf SZ thermal} & $-$ & large & y - distortion & thermal & [1] \\
 local group {\bf halo} & (Rayleigh--Jeans) & & & 1 keV & \\
 & & & & & \\
\hline
 & & & & & \\
 X-ray/radio & + & small & flat, $\, \alpha < 0.5$ & $\Gamma \approx
2.0 \pm 0.1$ & [2,3,4] \\
 point {\bf sources} & & & $(10 - 100 \, GHz)$ & $I = I_0
(E/E_0)^{-\Gamma}$ & \\
 & & & & & \\
\hline
 & & & & & \\
 {\bf ISW / RS} &  $+$ & large & Planck & $\Gamma \approx 2.0 \pm 0.1$ &
[5,6] \\
 in $\Lambda$ / open universe & & $\ell \approx 10$ & &  $I = I_0
(E/E_0)^{-\Gamma}$ & \\
 & & & & & \\
\hline
\end{tabular}
\caption{\normalsize Overview of different effects introducing 
possible correlations between microwave and X-ray data. Authors: 
[1] Suto et al. 1996, [2] Franceschini et al. 1989, [3] Franceschini
1995, [4] Laurent-Muehleisen et al. 1996, [5] Crittenden \& Turok
1996, [6] Kamionkowski 1996.}
\end{center}
\end{table*}

Comparison of the dipole components is of interest, because the 
firm detection of a cosmological X-ray dipole can either give confirmation 
to the velocity interpretation of the CMB dipole, or probe the
distribution of matter on intermediate scales (100 - 1000 h$^{-1}$
Mpc) and redshifts (z $<$ 5). The velocity dipole amplitude can be 
calculated from the Compton-Getting effect (Compton \& Getting 1935) 
incorporating the relativistic effects of aberration and spectral
shift to 
\begin{equation}
\frac{I_{CG}}{I} = (2+\Gamma) \frac{v}{c} \cos \Theta \, .
\end{equation}
The photon index (n(E) $\propto$ E$^{-\Gamma}$) for the extragalactic
XRB in the ROSAT energy bands is $\Gamma_{ROSAT} = 2.0 \pm 0.1$
(Hasinger et al. 1993) and the sun's 
velocity with respect to the CMB has been measured in the COBE DMR to 
be $v = 369.0 \pm 2.5$ km/s in the direction 
$(\ell,b) = (264^\circ.31 \pm 0^\circ.17,
+ 48^\circ.05 \pm 0.^\circ10)$ (Lineweaver et al. 1996). 
With these numbers, the expected dipole rms amplitude, expressed as a 
percentage of the monopole term, is D$_{CG} = 0.08$. 
This is more than an order of magnitude lower than our 
typically measured dipole amplitudes (cf. to the last column of table
1, which gives the observed values expressed in the same manner). 
A dipole resulting from the distribution of matter is theoretically 
less well determined, but is assumed to have comparable amplitude to
the Compton-Getting dipole (Lahav et al. 1996). 
Thus, the expected extragalactic dipoles have amplitudes lower 
than our observed values. 

Since we know from the energy dependence and orientation 
the Galactic origin of the fitted dipoles, our conclusion is 
that the ROSAT data over the energy range from 0.5 -- 2 keV are, 
in major regions of the sky, strongly dominated by a dipolar Galactic
structure and a considerable effort would be needed to separate out 
an extragalactic dipole, if existent in the data.

\subsection{Galactic Quadrupole}

Strong Galactic correlation between the data should not be expected. 
A spatial correlation between
regions of radio free-free (Bremsstrahlung) emission and H{\sc ii} regions
(H$_\alpha$) is plausible. However, it is still unclear if the X-ray signature
of H{\sc ii} regions is dominated by emission from associated hot gas
or absorption by associated neutral gas. Only in the first case would 
a positive correlation be expected.

Supernova remnants are 
prominent Galactic features in the soft X-rays, and enrich the Galactic 
medium with compressed magnetic fields that are responsible for synchrotron
emission also at microwave frequencies. However, under the usual 
assumptions regarding the spectral behaviour of these 
synchrotron sources, a strong contribution at COBE frequencies is not 
to be expected. This is in agreement with 
the lack of correlation (Kogut et al. 1996) between COBE DMR and 
the template for synchrotron radiation, the 408 MHz map (Haslam et al. 1982). 
The North Polar Spur, which traces the rim of the nearby superbubble 
Loop I (e.g. Egger et al. 1996), 
is however the most prominent synchrotron source
and one positive pole of the correlated quadrupole lies in the
corresponding direction. 
The amplitude of the correlated signal in 
the COBE DMR is not incompatible with the signal expected from this
object in the microwave band, given the uncertainty in the spectral index. 
Since the North Polar Spur is also known as a very bright X-ray 
source it is possible that this feature 
is associated with the correlation. 

Although the contribution of the correlated quadrupole 
to the CMB channels in the DMR is comparatively small, it will be interesting 
to investigate the meaning of the detection at 31.5 GHz further, by
comparison with the Galactic synchrotron radiation template and 
the template for 
Bremsstrahlung and dust radiation (which is the COBE DIRBE 140 $\mu$m
map). These topics will be addressed in a forthcoming paper. 
By doing this, a separation between 
synchrotron radiation and Bremsstrahlung might be possible. 
The answer to the question of whether a physical effect or 
purely geometrical alignment 
is responsible for the correlations could be decided. 

\subsection{Derivation of Limits}

On the selected NGP field some indication of a positive correlation 
of large angular scale was found. Since the correlated gradient in 
ROSAT is of Galactic origin, chance alignment with the COBE CMB
quadrupole, and possibly some spatial alignment with the COBE Galactic 
quadrupole (see section 5.2) turns out to be the cause, 
and not an extragalactic correlation. 
After subtracting the dipole and quadrupole terms from a Galactic cut
sky, or related large angular scale gradients from the selected 
NGP field, no significant correlations are found in the data. 
This will enables us to set limits on a number of possible mechanisms for 
introducing correlations as described in the introduction and summarized 
in table 2. We note here, that all our findings are in no conflict 
with the standard cosmological interpretation of the COBE DMR measurements. 

The large-scale, correlated feature we are seeing, the presumable
Galactic quadrupole, cannot originate from the SZ effect, as
would be the case for the Local Group halo, since the observed
correlation is positive. So at the quoted sensitivity, no Local Group
halo is found. By making use of the specific halo template the
sensitivity might be improved.

From the temperature limits on the selected NGP area we can
infer limits on the Comptonization
parameter $y$ on scales of 7$^\circ$--40$^\circ$ 
from distortions in the Rayleigh-Jeans regime 
\begin{equation}
\delta T / T = - 2 \delta y 
\end{equation}
and find $\delta y < 8 \times 10^{-7}$. More specific constrains will be
derived by combining the DMR and ROSAT limits with model assumptions 
about the distribution of the gas and its properties. For the gas halo 
model adopted by So\l tan et al. (1996a) as one explanation for the extended
correlated X-ray emission around clusters of galaxies, we can
limit the assumed temperature of smoothly distributed gas to less 
than 2 keV (95 \% CL). 
Miyaji et al. (1996), when correlating ROSAT to HEAO, found excess 
fluctuations at ROSAT energies in comparison to predictions from the 
population synthesis model by Comastri et al. (1995). These
fluctuations can be explained with excess emission by the gas halo
model only when a temperature of at least 2 keV is assumed. 
Combined with our limits, this leaves a very narrow, possible 
temperature range for the gas. 

Also the combined contribution to the microwave and X-ray fluxes 
of a population of radio-loud AGNs can be limited, taking into account 
that the sensitivity for the correlated locations is limited 
to the zero-lag convolved with the DMR beam.

No indications for a correlation through the ISW have been seen, which
should dominate in the high DMR frequencies and hard X-rays. 
In this case, where we are interested in correlations between the CMB
and the extragalactic XRB at a range of scales from a few degrees up
to the quadrupole, another search strategy could be adopted, 
correlating the selected NGP field to the DMR cut sky with the effect 
of slightly improving the sensitivity on scales of the size of the
field, as tests have shown, and quantifying the sensitivity on scales above. 
Since we do not know how well the fluctuations 
in the ROSAT XRB trace the projected gravitational potential, although
rough assumptions can be made from the properties of the resolved
sources (Comastri et al. 1995), we do not give any limits on 
a $\Lambda$-universe in this analysis.

Comparing the limits derived on the selected NGP field and on the
$30^\circ$ cut sky, we find them to differ only marginally. 
Two effects, reduction of $\sqrt{N}$-noise in COBE and increase of 
predominantly uncorrelated Galactic structure in ROSAT, when going from
the smaller to the larger area, can be seen. For
comparison with the extragalactic, diffuse XRB the limits from the
selected NGP field are preferred.

We qualitatively discussed how the limits we found constrain possible 
correlation mechanisms. A more quantitative study should make use of 
specific predictions by each individual effect for e.g. frequency and 
energy dependence, angular scale, spatial orientation etc. which
could result in stronger constraints. This analysis is in progress and
will be presented in a future publication.

On the observational side, great improvements for this kind of analysis
can be expected from future satellite experiments such as the CMB missions MAP
and COBRAS/SAMBA (Bersanelli et al. 1996), 
compared to high resolution X-ray observations such
as XMM and AXAF, and the hard X-ray survey ABRIXAS (Friedrich et
al. 1996). In terms of
statistical analysis a comparison between COBRAS/SAMBA and ABRIXAS 
appears most promising, since ABRIXAS will be the most sensitive
all-sky X-ray survey (0.3--10 keV) and COBRAS/SAMBA will be the 
CMB mission with highest angular resolution and best frequency
coverage, enabling a separation between different frequency
dependences such as Planck spectrum, 
$y$-distortions and radio source spectra to be achieved.

\acknowledgements
%________________________________________ Do not leave a blank line here!
RK is grateful to his supervisor G. B\"orner and thanks 
A.J. Banday, C.H. Lineweaver, S.D.M. White and the referee, O. Lahav,  
for helpful discussions and suggestions. The {\it ROSAT} project has
been supported by the Bundesministerium f\"ur Bildung, Wissenschaft,
Forschung und Technologie (BMBF/DARA) and by the Max-Planck-Society. The COBE
datasets were developed by the NASA Goddard Space Flight Center 
under the guidance of the COBE Science Working Group and were 
provided by the NSSDC.
%_____________________________________________________________________

\end{document}